\documentclass[conference]{IEEEtran}
\IEEEoverridecommandlockouts
\usepackage{cite}
\usepackage{amsmath,amssymb,amsfonts}
\usepackage{graphicx}
\usepackage{textcomp}
\usepackage{xcolor}
\usepackage{mathrsfs}
\usepackage{mathtools, cuted}
\usepackage{stmaryrd}
\SetSymbolFont{stmry}{bold}{U}{stmry}{m}{n}
\usepackage{lmodern}

\usepackage{tikz}

\usepackage{amsthm}
\usepackage{algorithm} 
\usepackage{algorithmic}

\usepackage{afterpage}

\usepackage{multicol}
\usepackage{float}
\usepackage{lipsum}
\floatstyle{plain}
\newfloat{twocolequfloat}{b}{zzz}
\floatname{twocolequfloat}{Equation}

\newcommand{\set}[1]{\mathcal #1} 
\renewcommand{\vec}[1]{\boldsymbol{#1}} 
\newcommand{\mat}[1]{\boldsymbol{\underline{#1}}} 
\newcommand{\prob}[2]{\mathbb{P}_{#2}\left[#1\right]} 
\newcommand{\Exp}[2]{\mathbb{E}_{#2}\left[#1\right]} 
\newcommand{\Cov}[2]{\mathbb{C}_{#2}\left[#1\right]} 
\newtheorem{theorem}{Theorem}
\newtheorem{lemma}{Lemma}

\newtheorem{definition}{Definition}
\DeclareMathOperator*{\argmin}{argmin}

\def\BibTeX{{\rm B\kern-.05em{\sc i\kern-.025em b}\kern-.08em
    T\kern-.1667em\lower.7ex\hbox{E}\kern-.125emX}}
\begin{document}

\title{Non-Asymptotic Achievable Rate-Distortion Region for Indirect Wyner-Ziv Source Coding}

\author{
 Jiahui Wei$^{1,2}$, Philippe Mary$^{2}$, and Elsa Dupraz$^{1}$ \\
\small $^{1}$ IMT Atlantique, CNRS UMR 6285, Lab-STICC, Brest, France \\ $^{2}$ Univ. Rennes, INSA, IETR, UMR CNRS, Rennes, France \thanks{ This work has received a French government support granted to the Cominlabs excellence laboratory and managed by the National Research Agency in the ``Investing for the Future'' program under reference ANR-10-LABX-07-01 and also by the program \emph{France 2030} under reference ANR-23-CMAS-0001. This work was also funded by the Brittany region.}
}

\maketitle

\tikzset{
box/.style ={
rectangle, 
rounded corners =5pt, 
minimum width =50pt, 
minimum height =20pt, 
inner sep=5pt, 
draw=black },
bigbox/.style ={
rectangle, 
minimum width =50pt, 
minimum height =65pt, 
inner sep=5pt, 
draw=black 
},
roundnode/.style={
circle, draw=black, 
very thick, 
minimum size=6mm},
squarenode/.style={
rectangle, 
draw=black, 
very thick, 
minimum size=7mm}
}

\begin{abstract}
In the Wyner-Ziv source coding problem, a source $X$ has to be encoded while the decoder has access to side information $Y$. This paper investigates the indirect setup, in which a latent source $S$, unobserved by both the encoder and the decoder, must also be reconstructed at the decoder. 
This scenario is increasingly relevant in the context of goal-oriented communications, where $S$ can represent semantic information obtained from $X$. 
This paper derives the indirect Wyner-Ziv rate-distortion function in asymptotic regime and provides an achievable region in finite block-length. 
Furthermore, a Blahut-Arimoto algorithm tailored for the indirect Wyner-Ziv setup, is proposed. This algorithm is then used to give a numerical evaluation of the achievable indirect rate-distortion region when $S$ is treated as a classification label.
\end{abstract}


\section{Introduction}
In this paper, we address distributed source coding, where data is collected remotely by sensors, compressed, and  transmitted to a distant server for further processing. We focus on a widely studied yet simplified scenario involving two sources, $X$ and $Y$, where $Y$ is directly available as side information to the decoder. When the objective is to reconstruct the source $X$ under a certain distortion constraint, this setup is referred to as Wyner-Ziv coding~\cite{wyner1976rate}.
Here, we consider an extension of the Wyner-Ziv problem, motivated by recent developments in goal-oriented communications where the focus shifts from data reconstruction to applying a specific task at the receiver. In our setup, illustrated in Figure~\ref{fig:ch7_semantic_wz}, the objective of the decoder includes recovering not only the observation $X$, but also a latent source $S$, which is not observed by either the encoder or the decoder.
The source coding problem is specified by two distortion constraints, one on the source $X$, and one on the source $S$, potentially defined from different distortion measures. For instance, in applications such as autonomous driving or video surveillance, accurate recovery of both a classification label $S$ (\emph{e.g.}, object type) and the original source $X$ (\emph{e.g.}, visual data) is crucial. 

This setup corresponds to an indirect source coding problem that was originally formalized in~\cite{dobrushin1962information,witsenhausen1980indirect}. These works derived the indirect rate-distortion function for a point-to-point scenario \textit{without} side information (in our work, the Wyner-Ziv setup is considered). Recent advancements have revisited the indirect rate-distortion problem in the context of goal-oriented communications, interpreting the source $S$ as the semantic information~\cite{liu2021rate}. The work~\cite{liu2021rate} also addressed the case of non-i.i.d. Gaussian sources, providing a method for the numerical evaluation of the corresponding rate-distortion function. Further extensions have been proposed for arbitrary finite alphabet sources~\cite{Stavrou2022} and in the conditional setup where the side information $Y$ is available at both the encoder and decoder~\cite{guo2022semantic}. The general multiterminal setup with $M$ encoders has also been explored recently in~\cite{tang2024distributed}. 

Prior works~\cite{dobrushin1962information,witsenhausen1980indirect,liu2021rate,Stavrou2022,guo2022semantic,tang2024distributed} primarily focused on the asymptotic regime, where the source sequence length $n$ tends to infinity. A notable exception is the recent work~\cite{yang2024joint}, which introduced a non-asymptotic analysis for the indirect point-to-point case \textit{without} side information. In this paper, as our main contribution, we investigate the non-asymptotic indirect Wyner-Ziv source coding problem, aiming to provide more practical insights compared to the asymptotic analysis.  

Analyzing finite blocklength regimes requires specific techniques. The Berry-Esseen Theorem, leading to the concept of dispersion~\cite{polyanskiy2010channel,kostina}, provides bounds onto the error made when approximating certain probability measures with a Gaussian one. While dispersion analysis has been applied to both channel~\cite{polyanskiy2010} and source coding problems~\cite{kostina}, extending this method to complex multiterminal problems remains a challenge. Recent works, such as~\cite{tan_three_nit}, derived the dispersion matrix for Slepian-Wolf coding using the entropy density vector, and~\cite{watanabe2015nonasymptotic} extended these methods to the Wyner-Ziv setup with channel resolvability codes. Additionally, the Poisson matching lemma introduced in~\cite{li2021poisson} has provided tighter bounds for multiterminal problems, including the Wyner-Ziv setup. Here, we extend this framework to consider both the observation $X$ and the latent source $S$. To the best of our knowledge, this work is the first to adapt the Poisson Matching Lemma to the indirect Wyner-Ziv coding problem. 

In addition, one important issue remains in the numerical evaluation of the indirect rate-distortion function for source models of interest. Therefore, in this paper, we provide a Blahut-Arimoto algorithm which allows us to numerically derive the optimal test channel and  evaluate the indirect rate-distortion function in the Wyner-Ziv setup at finite blocklength. Motivated by a goal-oriented problem in which the objective is to perform a classification task at the receiver, we further consider a binary source $S$ with a Hamming distortion constraint. We provide a numerical evaluation of the non-asymptotic achievable rate-distortion region for this model. 

The remainder of the paper is as follows. Section~\ref{problem_statement} defines the source model and the indirect Wyner-Ziv source coding scheme. Section~\ref{sec:asymptotic_analysis} presents the asymptotic rate-distortion analysis, while Section~\ref{sec:nonasymptotic} introduces the non-asymptotic analysis. Finally, Section~\ref{sec:numericalresults} provides numerical results for the example of classification.

\section{Problem statement}
\label{problem_statement}

\subsection{Notation}
Throughout this article, random variables and their realizations are denoted with capital and lower-case letters, respectively, \emph{e.g.}, $X$ and $x$. Random vectors of length $n$ are denoted $\vec{X} = \left[X_1, ..., X_n\right]^T$, and  
$\mathbb{E}[\vec{X}]$ and $\Cov{\vec{X}}{}$ are the expected value and the covariance matrix of $\vec{X}$, respectively. 
Sets are denoted with calligraphic fonts,  
and the indicator function is defined as $\mathbf{1}\left[x\in \set{A}\right] = 1$ if $x\in \set{A}$, and $0$ otherwise. Finally $\log(\cdot)$ is the base-2 logarithm.

We consider the measurable space $\left( \set{X} , \mathscr{B}\left(\set{X}\right) \right)$, where $\mathscr{B}\left(\set{X}\right)$ is the Borel $\sigma$-algebra on the set $\set{X}$. For two $\sigma$-finite measures $\nu$ and $\mu$ over $\set{X}$, we use $\nu \ll \mu$ to state that $\nu$ is absolutely continuous with respect to $\mu$. The probability measure $P_X$ over $\left( \set{X} , \mathscr{B}\left(\set{X}\right) \right)$ 
is the distribution of $X$. 
The notation $\prob{\cdot}{}$ is used for the probability of an event over the underlying probability space. For a joint probability measure $P_{XY}$ on $\set{X}\times \set{Y}$, the information density is denoted as~\cite{polyanskiy2010} 
\begin{equation}\label{eq:def_information_density}
    \iota\left(x,y\right) := \log \frac{d P_{Y\left|\right.X=x}}{d P_Y} \left(y\right),
\end{equation}
where the ratio above is the Radon-Nykodym derivative of the conditional measure $P_{Y\left|\right. X=x}$ with respect to the measure $P_Y$, in $y$.

\subsection{Source definitions}
\begin{figure}
   \centering

\begin{tikzpicture}
    \node (S)at (-1.5,-2) {$\vec{S}$};
    \node (X) at (0,-2) {$\vec{X}$};

    \node (hatX) at (7,-3) {$\hat{\vec{X}}$};
    \node (hatS) at (7,-1) {$\hat{\vec{S}}$};
    \node[squarenode,line width=0.8pt] (encoder) at (2,-2) { Encoder};
    \node[squarenode,line width=0.8pt] (decoder) at (4.5,-2) { Decoder};
    \node (Y) at (4.5, -3) {$\vec{Y}$};

    \draw[->,line width=0.6pt] (S) --  node[above] {$P_{\vec{XY}|\vec{S}}$} (X);
    \draw[->,line width=0.6pt] (-0.75, -2) -- (-0.75, -3)-- (Y);
    
    \draw[->,line width=0.6pt] (X) -- (encoder.west);
    \draw[->,line width=0.6pt] (encoder) -- (decoder.west);
    \draw[->,line width=0.6pt] (Y) -- (decoder.south);
    
    \draw[-,line width=0.6pt] (decoder) -- (6, -2);
    \draw[->,line width=0.6pt] (6, -2) -| (6,-3) -- (hatX);
    \draw[->,line width=0.6pt] (6, -2) -| (6,-1) -- (hatS);
    
    \draw[dashed,line width=0.5pt] (S) -- (-1.5, -0.25) --node[above] {$d_s(\vec{s}, \hat{\vec{s}})$} (7, -0.25) -- (hatS);
    \draw[dashed,line width=0.5pt] (X) -- (0, -3.75) --node[below] {$d(\vec{x}, \hat{\vec{x}})$} (7, -3.75) -- (hatX);
\end{tikzpicture}
    
    \caption{Coding scheme for goal-oriented Wyner-Ziv coding}
    \label{fig:ch7_semantic_wz}
\end{figure}
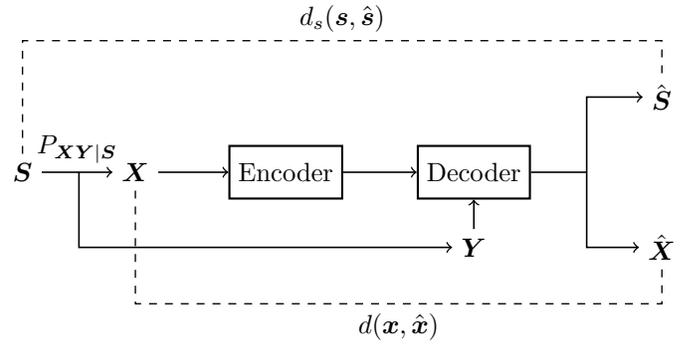

We consider length-$n$ source sequences $\vec{S}, \vec{X}, \vec{Y}$, where the triplet $(S_i,X_i,Y_i)$, $i\in \llbracket 1, n \rrbracket$, is i.i.d. and follows the distribution $P_{SXY} = P_{S} P_{XY|S}$ and takes values in the alphabet $\set{S} \times \set{X} \times \set{Y}$. 
As depicted in Figure~\ref{fig:ch7_semantic_wz}, the source sequence $\vec{X}$ is observed by the encoder while $\vec{Y}$ serves as side information available only at the decoder. The latent vector $\vec{S}$ is not observed by the encoder nor by the decoder. 

We consider two distortion measures 
\begin{align}
    d_s &: \set{S} \times \hat{\set{S}}  \rightarrow \mathbb{R}^+ \cup \{0\},\\
    d &: \set{X} \times \hat{\set{X}} \rightarrow \mathbb{R}^+ \cup \{0\},
\end{align}
where $d_s$ is the per-letter distortion measure on the latent source $S$, and $d$ is the per-letter distortion measure on the source $X$. In addition, $ \hat{\set{S}}$ and  $\hat{\set{X}}$ are the reconstruction alphabets for $S$ and $X$, respectively.  The distortions between the length-$n$ sequences are defined by:
\begin{align}
    &d_{n,s}(\vec{s}, \vec{\hat{s}}) = \frac{1}{n}\sum_{i=1}^{n}d_s(s_i, \hat{s}_i),\\
    &d_n(\vec{x}, \vec{\hat{x}}) = \frac{1}{n}\sum_{i=1}^{n}d(x_i, \hat{x}_i).
\end{align}

\subsection{Indirect Wyner-Ziv source coding scheme}
We now define indirect source coding schemes in asymptotic and non-asymptotic regimes. 
\begin{definition} [$(n, M_n, D, D_s)$-Wyner-Ziv code]
    An $(n, M_n, D, D_s)$-Wyner-Ziv code for the indirect source coding problem with $M_n = 2^{nR}$ consists of
    an encoding function 
    \begin{align}
    \label{eq_enc}
        \phi_n : \set{X}^n \rightarrow \{1, 2, ..., M_n\},
    \end{align}
    and a decoding function (possibly randomized)
    \begin{align}
    \label{eq_dec}
        g_n: \{1, 2, ..., M_n\} \times \set{Y}^n \rightarrow \hat{\set{S}}^n \times \hat{\set{X}}^n,
    \end{align}
    such that 
    \begin{align}\label{eq:dist_constraint_Ds}
        &\limsup_{n\rightarrow\infty} \Exp{d_{n,s}(\vec{S}, \vec{\hat{S}})}{}\leq D_s,\\
        &\limsup_{n\rightarrow\infty} \Exp{d_n(\vec{X}, \vec{\hat{X}})}{}\leq D. \label{eq:dist_constraint_D}
\end{align}
\end{definition}
In this definition, $R$ is the coding rate. The indirect Wyner-Ziv rate-distortion function is then given by
\begin{align}
    R_{WZ}(D, D_s) = \inf \{R : \ \exists \ \  (n, M_n, D, D_s)\text{-code} \} .
\end{align}

For finite $n$, some codewords may lead to distortions larger than the constraints $D$ and $D_s$. Therefore, we consider an excess probability $\epsilon$, and define an $(n, M, D, D_s, \varepsilon)$-code as follows. 
\begin{definition} [$(n, M_n, D, D_s, \varepsilon)$-Wyner-Ziv code]
    An $(n, M_n, D, D_s, \varepsilon)$-Wyner-Ziv code for indirect source coding with $M_n = 2^{nR}$ consists of
    an encoding function~\eqref{eq_enc} and a decoding function~\eqref{eq_dec} such that 
    \begin{align}
        \prob{(d_{n,s}(\vec{S}, \hat{\vec{S}}) \geq D_s) \cup
        (d_n(\vec{X}, \hat{\vec{X}}) \geq D) }{} \leq \varepsilon.
\end{align}
\end{definition}
The finite blocklength indirect Wyner-Ziv rate-distortion function with excess probability $\epsilon$ is
\begin{align}\label{eq:rd_secondorder} \notag
    R_{WZ}(n, D, D_s, \varepsilon) = \inf \{R : \ \exists \ \  (n, M_n, D, D_s, \varepsilon)\text{-code} \}. 
\end{align}


\section{Asymptotic Rate-distortion analysis}\label{sec:asymptotic_analysis}
In this section, we extend the conventional Wyner-Ziv rate distortion function to the indirect case in the asymptotic regime. We first provide the rate-distortion function, and then investigate its properties. 

\subsection{Indirect Wyner-Ziv rate-distortion function}
\begin{theorem}[Indirect Wyner-Ziv rate-distortion function]
\label{thm_swz}
    The rate-distortion function for indirect source coding with side information at the decoder and i.i.d. sources is given as the solution to the following optimization problem: 
    \begin{equation}
    \label{eq_rd_wz}
        R_{\text{WZ}}\left( D, D_s \right) = \hspace{-1cm} \mathop {\inf }\limits_{
        \begin{array}{c} P_{U\mid X}: \\ \Exp{ d(X,\hat{X}) }{} \leq D \\  \Exp{ d'(X,\hat{S}) }{} \leq D_s
        \end{array}} 
        \hspace{-1cm} I(X; U) - I(U; Y) ,
    \end{equation}
    given that the Markov chain $U -X- Y$ holds, and given the existence of a  mapping $g_n: \set{U} \times \set{Y}^n \rightarrow \hat{\set{S}}^n \times \hat{\set{X}}^n$ such that the distortion constraints $D$ and $D_s$ are satisfied. 
   We further denote $\hat{s} = \hat{s}(u, y)$ and $ \hat{x} = \hat{x}(u, y)$ when referring to the reconstruction of $\hat{s}$ or $\hat{x}$. 
    The modified distortion measure $d'(x, \hat{s})$ for the indirect source is defined by:
    \begin{align}
    \label{eq_d_s}
        d'(x, \hat{s}) &= d'(x, \hat{s}(u, y))  \notag\\
        &= \frac{1}{P_{XY}(x, y)}\sum_{s\in \set{S}} P_{XYS}(x, y, s)d_s(s, \hat{s}(u, y)).
    \end{align}
\end{theorem}
The proof of Theorem \ref{thm_swz} follows from standard achievability and converse proofs~\cite{thomas2006elements} for Wyner-Ziv problem. The modified distortion measure $d'$ in~\eqref{eq_d_s} comes from~\cite{witsenhausen1980indirect}. This result extends the indirect source coding problem without side information from~\cite{liu2021rate,Stavrou2022}, and is proved in Appendix \ref{apx_d:thm_swz}.


\subsection{Properties of the indirect Wyner-Ziv rate-distortion function}

The following results extend those of the conventional Wyner-Ziv rate-distortion problem~\cite{wyner1976rate, wyner1978}, and of the indirect rate-distortion without side information from~\cite{Stavrou2022}.
\begin{lemma}
\label{lemma_convex}
The indirect Wyner-Ziv rate-distortion function $R_{WZ}(D, D_s)$ is non-increasing and convex in $(D, D_s)$.    
\end{lemma}

\begin{lemma}
\label{lemma_bound_swz}
    The rate-distortion function $R_{WZ}(D, D_s)$ satisfies
    \begin{align}
    \label{eq_bound_wz}
        &\max \{R_{\text{WZ}}(D, \infty), R_{\text{WZ}}(\infty, D_s)\} \leq R_{\text{WZ}}(D, D_s) \notag\\
        &\quad\leq R_{\text{WZ}}(D, \infty) + R_{\text{WZ}}(\infty, D_s),
    \end{align}
    where $R_{\text{WZ}}(D, \infty), R_{\text{WZ}}(\infty, D_s)$, are the conventional Wyner-Ziv rate-distortion functions 
    for the reconstruction of $X$ only and for the reconstruction of $S$ only, respectively.
\end{lemma}
\begin{proof}
The proof of Lemma~\ref{lemma_convex} follows the same steps as for the conventional Wyner-Ziv rate-distortion function in~\cite{wyner1976rate}, and the details are provided in Appendix~\ref{apx_d:lemma_convex}.    
\end{proof}

\subsection{Optimal test channel by Blahut-Arimoto algorithm} 


\label{subsec_ba}

In the conventional Wyner-Ziv problem  which does not consider the latent source $S$, closed-form expressions of the rate-distortion function are only available for some specific source models, \emph{i.e.}, Gaussian sources with squared error distortion or discrete sources with Hamming distance. For other models, the rate-distortion function is not available in closed-form but can be evaluated using the Blahut-Arimoto (BA) algorithm \cite{arimoto1972algorithm}. The authors in~\cite{Stavrou2022} proposed a generalized version of the BA algorithm for indirect source coding without side information. In this section, we extend the work of~\cite{Stavrou2022} to propose BA algorithm for the indirect Wyner-Ziv setup.  

The Lagrangian associated with the optimization problem stated in Theorem~\ref{thm_swz} can be expressed as:
\begin{align}
\label{eq_lagrange}
    &\set{L}\left(P_{U|X}(u|x), \lambda, \mu, \xi(x)\right) = I(X; U) - I(U; Y) \notag\\
    &+ \lambda \left( \Exp{d(X, \hat{X})}{} - D\right)  + \mu \left(\Exp{d'(X, \hat{S})}{}  - D_s\right) \notag\\
    &+ \sum_x\xi(x) \left(\sum_u P_{U|X}(u|x) - 1\right),
\end{align}
where $\lambda\geq 0, \mu\geq0$ are the Lagrange multipliers corresponding to the distortion constraints, which can be stated per-letter due to the i.i.d. hypothesis, and $\xi(x)\geq 0$ is required to enforce $P_{U|X}$ to be a distribution.
By setting the differentiation of \eqref{eq_lagrange} with respect to $P_{U|X}$ to zero, we obtain the optimal conditional distribution $P_{U|X}^\star$ given in~\eqref{eq_optimal_p}, on the top of the next page, for the indirect Wyner-Ziv rate-distortion function~\eqref{eq_rd_wz}.

\begin{figure*}[!t]
\begin{equation}
        \label{eq_optimal_p}
            P_{U|X}^\star(u|x) = \frac{\exp \left ({\sum_y P_{Y|X}(y|x)\left( \log P_{U|Y}(u|y) - \lambda d(x, \hat{x}) - \mu d'(x, \hat{s} )\right)}\right)}{\sum_u \exp \left ({\sum_y P_{Y|X}(y|x)\left( \log P_{U|Y}(u|y) - \lambda d(x, \hat{x}) - \mu d'(x, \hat{s} )\right)}\right)}.
    \end{equation}
    \hrulefill 
\end{figure*}

We extend the BA algorithm for conventional Wyner-Ziv rate-distortion function~\cite{willems1983computation} to the indirect case provided in Theorem~\ref{thm_swz}. 
The algorithm takes as input some initial values for the Lagrange multipliers $\lambda$ and $\mu$, and a maximum number of iterations $N$. The core of the algorithm lies in updating at each iteration $\ell \in \llbracket 1,N\rrbracket$ the conditional distribution $P^{(\ell)}_{U|X}$ using~\eqref{eq_optimal_p}, as well as a reconstruction function $g^{(\ell)}:\set{U} \times \set{Y}^n \rightarrow \hat{\set{S}}^n \times \hat{\set{X}^n}$, where $|\set{U}| = |\set{X}| + 1$ and
\begin{align}
\label{eq_update_g}
    &g^{(\ell)}(u, y) = \underset{\hat{x}, \hat{s}}{\arg \min} \ \sum_x P^{(\ell-1)}_{U|X}(u|x)\left(\lambda  d(x, \hat{x}) + \mu d'(x, \hat{s}) \right) 
    \notag\\ 
    &\quad\text{for} \  \hat{x} \in \hat{\set{X}}, \hat{s} \in \hat{\set{S}}, \text{and} \  \ell\in \llbracket1,N\rrbracket. 
\end{align}
The algorithm stops when the maximum number of iteration is reached, or if the difference in distortions between two successive iterations is below a threshold $\delta$. By varying the values of $\lambda$ and $\mu$, we can control the trade-off between the distortions achieved on $X$ and $S$ and obtain the corresponding indirect rate distortion-function.

\section{Non-asymptotic achievability bound}\label{sec:nonasymptotic}

\subsection{Conditional Poisson Matching Lemma}
We first present the Poisson functional representation provided in~\cite{li2017functional}.  
%
Let $\set{U}$ be a Polish space with its Borel $\sigma$-algebra and a base measure $\nu$. Given that $\Lambda$ is the Lebesgue measure over $\mathbb{R}^+$, a Poisson process generates random variables $\{U_i, T_i\}_{i\in \mathbb{N}}$ on $\set{U} \times \mathbb{R}^+$, with intensity measure $\nu \times \Lambda$. For a measure $P \ll \nu$ on $\set{U}$, the objective is to construct a random variable $\Tilde{U}$ that has the distribution $P$ using the points of the Poisson process. The representation $\Tilde{U}_k$ is chosen such that $k \triangleq \argmin_i \frac{T_i}{(dP/d\nu)(U_i)}$ when $\frac{dP}{d\nu}(U_i)>0$. 


\begin{lemma}[Conditional Poisson Matching Lemma~\cite{li2021poisson}]
    For a joint distribution $P_{UXY}$ and two conditional distribution $P_{U|X}$ and $P_{U|Y}$ on Polish space $\set{U}$, with $P_{U|X}, P_{U|Y} \ll \nu$ almost surely. The random variables $\Tilde{U}_{P_{U|X}}$ and $\Tilde{U}_{P_{U|Y}}$ are generated according to $P_{U|X}$ and $P_{U|Y}$ respectively and we denote $U = \Tilde{U}_{P_{U|X}}$. We have
    \begin{align}
    \label{eq_poisson_matching_lm}
        \prob{\left.\Tilde{U}_{P_{U|Y}} \neq U \right| UXY}{} \leq 1 - \left( 1+ \frac{d P_{U|X}(\cdot|X)}{d P_{U|Y}(\cdot|Y)}(U)\right) ^{-1} .
    \end{align}
\end{lemma}
The mapping $P_{U|X}$ is used at the encoder, while the mapping $P_{U|Y}$ is used at the decoder. 
The previous lemma was used in~\cite[Theorem~3]{li2021poisson} to provide an upper bound on the excess probability for the conventional Wyner-Ziv problem. 

\subsection{Bound on the excess probability}
The following Theorem provides a bound on the excess probability in the indirect Wyner-Ziv source coding problem. 
\begin{theorem}
\label{thm_excess}
    For any fixed $P_{U|X}$ and reconstruction function $g_n: \set{U}^n\times\set{Y}^n \rightarrow \hat{\set{S}}^n  \times \hat{\set{X}}^n$, there exists an $(n,M_n,D, D_s, \varepsilon)$-code for the indirect Wyner-Ziv source coding problem such that 
    \begin{align}
        \varepsilon \leq &\mathbb{E}\left[1 - \mathbf{1}\left[d_s(S, \hat{S}) \le D_s \cap
        d(X, \hat{X}) \le D\right]\right. \notag\\ \label{eq:bound_excess}
        &\quad \left. \times\left(1 + M_n^{-1} 2^{\iota(U;X) - \iota(U;Y)}\right)^{-1} \right],
    \end{align}
    given that $P_{UX}\ll P_U \times P_X$ and $P_{UY}\ll P_U \times P_Y$.
\end{theorem}
\begin{proof}
    Let $\{U_i, L_i, T_i\}_{i\in \mathbb{N}}$ be the points of a Poisson process with intensity measure $P_X\times P_L \times \Lambda$, where $P_L$ is a uniform distribution over $\llbracket1,M_n\rrbracket$.
    Consider the following coding scheme. Upon observing $X\sim P_X$, the encoder generates a message $m \in \llbracket1,M_n\rrbracket$ based on the conditional distribution $P_{U|X}$ using the Poisson functional representation. The decoder receives the message $m$ and by a Poisson functional representation using the distribution $P_{U|Y}$, it outputs the reconstructed sources as $\hat{S}, \hat{X}= g(\Tilde{U}_{P_{U|Y}}, Y)$. Defining $U = \Tilde{U}_{P_{U|X}\times P_L}$, we have
    \begin{align}
        &\quad \prob{d_s(S, \hat{S}) \geq D_s \cup
        d(X, \hat{X}) \geq D }{}\\
        &\le 1 - \prob{d_s(S, \hat{S}) \le D_s \cap
        d(X, \hat{X}) \le D \cap \Tilde{U}_{P_{U|X}} = \Tilde{U}_{P_{U|Y}}}{} \notag\\ 
        &\le  \mathbb{E}_{P_L P_{UXY}}\left[1 -\mathbf{1}\left[d_s(S, \hat{S}) \le D_s \cap
        d(X, \hat{X}) \le D\right] \right .\notag\\
        & \quad \left . \times\prob{\left. \Tilde{U}_{P_{U|Y}} = U  \right | UXY}{}\right]\\
        &\le  \mathbb{E}\left[1 -\mathbf{1}\left[d_s(S, \hat{S}) \le D_s \cap
        d(X, \hat{X}) \le D\right] \right .\notag\\
        & \quad \left . \times \left( 1+ \frac{d \left(P_{U|X}\left(\cdot\left|\right. X\right) \times P_L \right)}{ d P_{U|Y}\left(\cdot\left|\right. Y\right)}(U) \right)^{-1} \right]\label{eq_pf_cpm}\\
        &\le  \mathbb{E}\left[1 -\mathbf{1}\left[d_s(S, \hat{S}) \le D_s \cap
        d(X, \hat{X}) \le D\right] \right .\notag\\
        & \quad \left . \times \left( 1 + M_n^{-1} 2^{\iota(U;X) - \iota(U;Y)} \right)^{-1} \right],
    \end{align}
    where~\eqref{eq_pf_cpm} comes from the conditional Poisson matching lemma \eqref{eq_poisson_matching_lm} on $P_{U|X}$ and $P_{U|Y}$.
\end{proof}
Unlike the non-asymptotic proofs based on packing and covering lemmas~\cite{verdu,watanabe2015nonasymptotic}, 
the Poisson matching lemma simultaneously bounds the information densities $\iota(U;X)$ and $\iota(U;Y)$ which results in a tighter non-asymptotic achievability bounds~\eqref{eq:bound_excess} than the one that would be obtained using the channel resolvability codes~\cite{watanabe2015nonasymptotic}, as shown in~\cite{li2021poisson} for the Wyner-Ziv problem.

\subsection{Second-order rate-distortion region}
We now provide the second-order achievable rate-distortion region for the indirect Wyner-Ziv setup. 
By using the condition $\iota(U;X) - \iota(U;Y) > \gamma$ for any positive $\gamma$, \eqref{eq:bound_excess} can be rewritten as
\begin{align}
\label{eq_excess_bound}
    \varepsilon &\le \mathbb{P}\left[\iota(U;X) - \iota(U;Y) > \gamma \cup  d_s(S, \hat{S}) > D_s\right.\notag\\
    &\qquad \ \left. \cup\  d(X, \hat{X}) > D \right] + \frac{2^{\gamma}}{M}.
\end{align}
We define the positive semi-definite matrix $\mat{V}\in\mathcal{R}^{3\times 3}$ as
$$\mat{V} = \Cov{\left[\iota(U;X) - \iota(U;Y), d(X, \hat{X}) , d_s(S, \hat{S})\right]^T}{} \in \mathbb{R}^{3\times 3} .$$  
For the Gaussian random vector $\mathbf{B}\sim \mathcal{N}(0,\mat{V})$, the dispersion region is defined as~\cite{tan_three_nit}
\begin{align}
\label{dispersion}
    \mathscr {S}( \mat {V}, \varepsilon ):= \{ \mathbf {b}\in \mathbb {R}^{3}: \prob{\mathbf {B}\le \mathbf {b}}{}\ge 1- \varepsilon \}.
\end{align}
%
Let $\gamma = \log M - \log n$. By using the multidimensional Berry-Esséen Theorem as in \cite{watanabe2015nonasymptotic, jwei2024distributed}, we provide the second-order indirect Wyner-Ziv rate-distortion function as follows. 
\begin{theorem}
\label{thm_second_coding_rate}
    Let $\vec{e}_1 = [1\quad 0\quad 0]$, for any $0<\varepsilon<1$ and $n$ sufficiently large, we have 
    \begin{align}
        &R_{WZ}(n, D, D_s, \varepsilon) \notag\\
        \le& \inf_R \left\{R_{WZ}(D, D_s) + \vec{e}_1\frac{\mathscr{S}(\mat{V}, \varepsilon)}{\sqrt{n}} + 2\frac{\log n}{n}\right\} .
    \end{align}
\end{theorem}
\begin{proof}
    The proof is provided in Appendix~\ref{apx:second_order_coding_rate}.
\end{proof}

\vspace{-0.2cm}

 We observe that the second-order rate-distortion function depends on the asymptotic function $R_{WZ}(D, D_s)$, penalized by two terms that vanish with $n$. The infinum is taken on the rate $R$ that implicitly appears in the second term (cf Appendix \ref{apx:second_order_coding_rate}).



\section{Example: classification}\label{sec:numericalresults}

We now focus on a classification problem where the latent variable $S$ represents the class label. In this setup, the correlation between $X$ and $Y$ is modeled by a Gaussian mixture, with $S$ parameterizing the mixture component. 

\subsection{Source Model}
Let us assume a binary source $S \in \{0, 1\}$, uniformly distributed, and a joint Gaussian pair $(X,Y)$ with
\begin{align}
    (X, Y)\sim \mathcal{N}(\vec{0}, \mat{\Sigma_0}) \quad &\text{if}\quad S = 0,\\
    (X, Y) \sim \mathcal{N}(\vec{0}, \mat{\Sigma_1}) \quad &\text{if}\quad S = 1.
\end{align}
The covariance matrices $\mat{\Sigma_0}, \mat{\Sigma_1}$, are defined as follows: 
\begin{align}
\label{eq_source_gaus_xy}
    \mat{\Sigma_0} = \begin{bmatrix}
    \sigma_X^2 & \theta_0 \\
    \theta_0 & \sigma_Y^2 
    \end{bmatrix}, \ \text{and} \ \mat{\Sigma_1} = \begin{bmatrix}
    \sigma_X^2 & \theta_1\\
    \theta_1 & \sigma_Y^2 
    \end{bmatrix},
\end{align}
where $\theta_0, \theta_1 \in \mathbb{R}$ are the covariance between $X$ and $Y$,  such that the matrices $\mat{\Sigma_0}$ and $\mat{\Sigma_1}$ are positive semi-definite. 

In this setup, the state $S_i$ should be estimated for every pair $(X_i, Y_i)$, $i\in \llbracket 1, n\rrbracket$. It cannot be inferred from a single observation, \emph{e.g.}, only $X_i$ or only $Y_i$, but requires joint observation of the pair $(X_i, Y_i)$. Therefore, $S$ cannot be estimated at the encoder. Hence a compressed version of $X$ should be transmitted in order to make the classification at the decoder.
Considering the Hamming distance for the semantic distortion $d_s$ and the mean squared error for the observation distortion $d$, this problem corresponds to the investigation of a rate-distortion-classification function and allows us to investigate the trade-off between rate, distortion, and classification accuracy.


\subsection{Numerical results} \label{sec:simuls}
\begin{figure}[t]
    \centering
    
        \includegraphics[width=\linewidth]{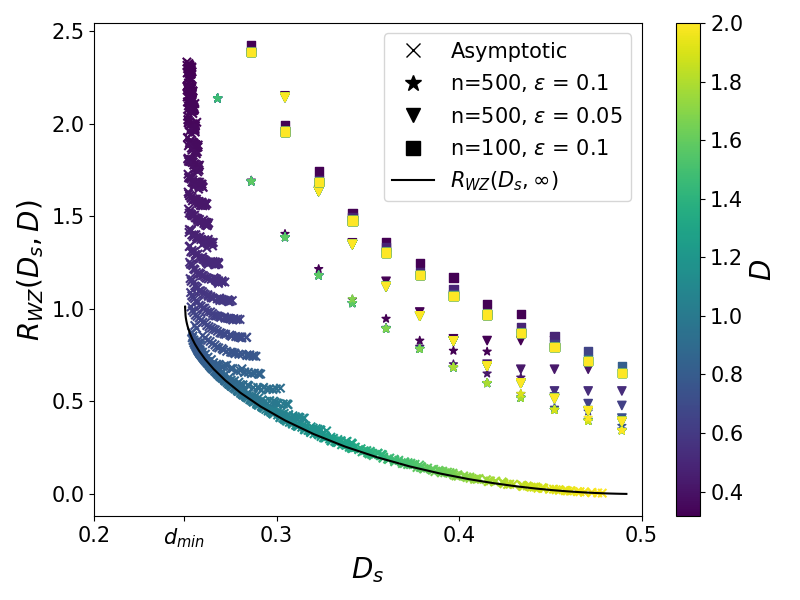}
        \vspace{-0.5cm}
        \label{fig:rdc_2d}
    \caption{$R_{WZ}(D, D_s)$ in finite
block-length regimes.}
    \label{fig:rdc}
\end{figure}


Given that with the previous model,  the optimal test channel $P_{U|X}$ is unknown and difficult to be derived analytically, we propose to numerically evaluate the second-order coding rate as follows. First, the optimal test channel $P^\star_{U|X}$ is  obtained by BA algorithm introduced in Section~\ref{subsec_ba}. Based on this test channel, we obtain the asymptotic coding rate $R_{WZ}(D, D_s)$. Finally, the dispersion region is obtained by evaluating the covariance matrix $\mat{V}$ for the considered blocklength $n$.

In the sequel, we set $\sigma_X^2 = 2, \sigma_Y^2 = 1, \theta_0 = 1, \theta_1 = -1$. We use 100 discretization levels between $-10$ and $10$, for both $X$ and $Y$, in order to solve~\eqref{eq_update_g} by exhaustive search.
Furthermore, the maximum iteration and tolerance are set to $N=100$ and $\delta = 0.001$, respectively. Figure~\ref{fig:rdc} illustrates the achievable region in the finite blocklength regime, with distortion on $X$ indicated by the color bar, for various values of $n$ and $\varepsilon$. The asymptotic region is also depicted for comparison.
We observe that as $n$ or $\varepsilon$  increases, the achievable region in the non-asymptotic regime expands towards the asymptotic region. This highlights the trade-off between blocklength, excess probability, and distortion.




Additionally, when $n$ is small,~\emph{e.g.}, $n=100$, the achievable bound is loose. Indeed, according to the bound, a non-zero rate is required to achieve a semantic distortion $D_s$ greater than $0.5$. However, this distortion can be trivially achieved with zero rate through random guessing at the receiver, even in the finite blocklength regime. This discrepancy arises from the Gaussian approximation, which introduces a constant penalty of order $O(\frac{\log n}{n})$.  However, for larger values of $n$ we observe that the rate approaches zero as $D_s$ approaches $0.5$.




\section{Conclusion} \label{conclusion}
In this paper, we investigated asymptotic and non-asymptotic indirect Wyner-Ziv source coding, with distortion constraints on both the observation and the latent source. 
By applying the Poisson Matching lemma, we derived the non-asymptotic achievability bound and determined the second-order coding rate. 
We also proposed a modified BA algorithm to numerically evaluate the indirect rate-distortion function. Future works will focus on more complex machine learning tasks, such as clustering or classification with more than two classes. 

\bibliographystyle{IEEEtran}
\bibliography{draft}

\appendices
\section{Proof of Theorem \ref{thm_swz}}
\label{apx_d:thm_swz}

The proof is a combination of the rate-distortion problem with two constraints \cite{thomas2006elements} and the conventional Wyner-Ziv \cite{wyner1976rate} coding problem.

\subsection*{Achievability}

The achievability part is a straightforward extension of the joint typicality test coding with binning for lossy source coding with side information at the decoder \cite{wyner1976rate}, when an additional constraint on the distortion of $S$ is considered. Fix the conditional test channel $P_{U|X}$ and further consider the reconstruction function $g : \set{U} \times \set{Y} \rightarrow \hat{\set{S}} \times \hat{\set{X}}$. 

\textbf{Generation of codebook: } The codewords are generated following the distribution $P_U$. First generate $2^{nR_1}$ $i.i.d.$ codewords following the output distribution of test channel $P_{U|X}$, \emph{i.e.} $i \in \{1, 2, ... , 2^{nR_1}\}$. Randomly and independently reassign these indices into $2^{nR_2}$ bins, with $B(m)$ denoting the indices assigned to bin $m$. We have $R_1 = I(X; U)$ and $R_2 = I(X; U) - I(U; Y)$, and there are averagely $2^{n(R_1-R_2)}$ indices per bin. The whole codebook $\mathcal{C}$ is revealed to both encoder and decoder.

\textbf{Encoding: }Given a sequence $\vec{x} \in \set{X}^n$, the encoder finds a codeword $\vec{u}(i)$ such that $\vec{u}(i)$ and $\vec{x}$ are jointly typical. If there is no such codeword, the encoder sets $i=1$. The encoder sets $i$ to the smallest one if there are more than one such codewords. Then the encoder sends the index of bins $m$ in which the codeword $i$ belongs.

\textbf{Decoding: }Upon receiving the index $m$, the decoder looks for a codeword $i$ in the bin $m$ such that $\vec{u}(i)$ and $\vec{y}$ are joint typical. It outputs the function $g(\vec{u}(i), \vec{y})$ if a unique $i$ is found. If no such $i$ or more than one $i$ is found, the decoder takes an arbitrary index in this bin.

\textbf{Analysis of error probability: }The following error events can be considered:
\begin{itemize}
    \item $\mathcal{E}_1$: The pair $(\vec{x},\vec{y})\notin \mathcal{A}_\epsilon^{(n)}$where $\mathcal{A}_\epsilon^{(n)}$ denotes the jointly typical set~\cite[Chapter~14]{thomas2006elements}. This probability tends to zero as $n\rightarrow\infty$ by the law of large numbers.
    \item $\mathcal{E}_2$: $\vec{x}$ is typical but there is no codeword $m$ such that $(\vec{x}, \vec{u}(i))$ are jointly typical, this probability goes to $0$ as long as $n$ is large enough and $R_1> I(X; U)$.
    \item $\mathcal{E}_3$: $(\vec{x}, \vec{u}(i))$ is jointly typical but $(\vec{u}(i), \vec{y})$ is not. This probability goes to zero by Markov lemma for $n$ large enough~\cite[Lemma~14.8.1]{thomas2006elements}.
    \item $\mathcal{E}_4$: There is another sequence $i'\neq i$ in the same bin such that $(\vec{u}(i'), \vec{y})$ is jointly typical. Without loss of generality, suppose bin index $m=1$ is received by the decoder : 
    \begin{align}
        &\prob{\mathcal{E}_4}{} \notag
        \\&= \sum_{\vec{u}, \vec{y}} P_{\vec{U}\vec{Y}}(\vec{u}, \vec{y})\prob{\exists i'\neq i, (\vec{u}(i'), \vec{y})\in\mathcal{A}_\epsilon^{(n)}(U, Y)}{}\\
        &= \sum_{\vec{u}, \vec{y}} P_{\vec{U}\vec{Y}}(\vec{u}, \vec{y}) \sum_{i'\neq i: (\vec{u}(i'), \vec{y})\in\mathcal{A}_\epsilon^{(n)}(U, Y)} \prob{i'\in B(1)}{}\\
        &\leq \sum_{\vec{u}, \vec{y}} P_{\vec{U}\vec{Y}}(\vec{u}, \vec{y}) \sum_{i': (\vec{u}(i'), \vec{y})\in\mathcal{A}_\epsilon^{(n)}(U, Y)} 2^{-n(R_1 - R_2)}\label{eq_pf_pb_bin}\\
        &\leq \sum_{\vec{u}, \vec{y}} P_{\vec{U}\vec{Y}}(\vec{u}, \vec{y}) 2^{n(I(U; Y) + \epsilon) } 2^{-n(R_1 - R_2)}\label{eq_pf_bd_jts}\\
        &\leq 2^{-n(R_1 - R_2 - I(U; Y) - \epsilon)},
    \end{align}
    where \eqref{eq_pf_pb_bin} is because the $2^{nR_1}$ indexes are independently and randomly assigned to $2^{nR_2}$ bins, \eqref{eq_pf_bd_jts} follows from a property of the joint typical set, i.e. $|\mathcal{A}_\epsilon^{(n)}(U, Y)| \leq 2^{n(I(U; Y) + \epsilon)}$ and $\epsilon\rightarrow0$ as $n\rightarrow\infty$. It suffices to take $R_1 - R_2 < I(U; Y) +  \epsilon$ to make this probability tend to zero as $n  \rightarrow \infty$.

    So $\prob{\mathcal{E}}{} \leq \sum_{i=1}^4\prob{\mathcal{E}_i}{}$ tends to 0 as $n  \rightarrow \infty$. By then we are sure to be able to find $(\vec{\hat{X}}, \vec{\hat{S}})$ that achieves the distortion constraint $(D, D_s)$
\end{itemize}

\subsection*{Converse}

We briefly restate the main step for the converse. It follows the proof for the converse of conventional Wyner-Ziv coding in \cite{wyner1976rate, thomas2006elements}. Denote $T$ as the output of encoder and define $U_i = (T, \vec{Y}^{i-1}, \vec{Y}_{i+1}^n)$ which satisfies the Markov relation $U-X-Y$. In the sequel, $\vec{Y}_i^j$, $1<i<j<n$ denotes a vector made of the elements of $\vec{Y}$ taken from the indices $i$ to $j$. We have:
\begin{align}
    nR &\geq H(T)\\
    &\geq H(T|\vec{Y})\\
    &\geq I(\vec{X}; T|\vec{Y})\\
    &=\sum_{i=1}^n I(X_i; T|\vec{Y}, X^{i-1})\label{eq_apxd_pf_chain}\\
    &=\sum_{i=1}^n H(X_i|Y_i) - H(X_i|T, \vec{Y}, \vec{X}^{i-1})\\
    &\geq \sum_{i=1}^n H(X_i|Y_i) - H(X_i|T, \vec{Y}^{i-1}, Y_i,  \vec{Y}_{i+1}^n)\\
    &=\sum_{i=1}^n H(X_i|Y_i) - H(X_i|U_i, Y_i)\label{eq_apxd_pf_defu}\\
    &=\sum_{i=1}^n I(X_i; U_i) - I(U_i; Y_i)\\
    &\geq \sum_{i=1}^n R_{WZ}\left(\Exp{d(X_i, \hat{X}_i)}{}, \Exp{d'(X_i, \hat{S}_i)}{}\right)\label{eq_apxd_pf_wz}\\
    &\geq nR_{WZ} \left( \Exp{d_n(\vec{X}, \hat{\vec{X}})}{}, \Exp{d'_n(\vec{X}, \hat{\vec{S}})}{}\right)\label{eq_apxd_pf_jensen}\\
    &=nR_{WZ}(D, D_s),
\end{align}
where \eqref{eq_apxd_pf_chain} follows from the chain rule for mutual information, \eqref{eq_apxd_pf_defu} is from the definition of $U_i$, \eqref{eq_apxd_pf_wz} holds since the \emph{information} rate-distortion function is such that $I(X; U) - I(U;Y) \geq \inf_{P_{U|X}: \Exp{d(X, \hat{X})}{}\leq D, \Exp{d'(X, \hat{S})}{}\leq D_s}I(X; U) - I(U;Y)$, and \eqref{eq_apxd_pf_jensen} is because of Jensen's inequality and the convexity of $R_{WZ}(\cdot, \cdot)$.

\section{Proof of Lemma \ref{lemma_convex} and Lemma~\ref{lemma_bound_swz}}
\label{apx_d:lemma_convex}
We now show that the Wyner-Ziv rate-distortion function $R_{WZ}(D, D_s)$ is both non-increasing and convex in $(D, D_s)$.
\paragraph{Non-increasing: } 
The monotonicity of $R_{WZ}(D, D_s)$ follows from the fact that, as $D$ or $D_s$ increases, the minimization set that satisfies the distortion constraints in \eqref{eq_rd_wz} becomes larger. Since $R_{WZ}(D, D_s)$ is the minimum of mutual information $I(X; U) - I(U; Y)$ over this set, it must be non-increasing in $(D, D_s)$.

\paragraph{Convex: }
Let $(D_1, D_{s1})$ and $(D_2, D_{s2})$ be two pairs of distortion values, and let $(U_1, g_1)$, $(U_2, g_2)$ be the corresponding auxiliary random variables and reconstruction functions that achieve the minimum rate-distortion function $R_{WZ}(D_1, D_{s1})$ and $R_{WZ}(D_2, D_{s2})$, respectively. 

We now consider a new random variable $U_3$ by taking $U_1$ with probability $\lambda$ and $U_2$ with probability $1-\lambda$, where $\lambda \in [0,1]$. Similarly, let the reconstruction function $g_3(U_3, Y)$ be defined as 
\begin{align}
    g_3(U_3, Y) = 
    \begin{cases}
        g_1(U_1, Y) \quad &\text{with probability }\lambda, \\
        g_2(U_2, Y), & \text{with probability }1-\lambda.
    \end{cases}
\end{align}
Define $\hat{X} = g(U, Y)_1, \hat{S} = g(U, Y)_0$. The corresponding distortions $D_3$ and $D_{s3}$ for $U_3$ can be expressed as:
\begin{align}
    \begin{bmatrix} D_3 \\ D_{s3} 
    \end{bmatrix} 
    &= \lambda \begin{bmatrix} \Exp{d(X, g_1(U_1, Y)_1)}{}\\ \Exp{d'(X, g_1(U_1, Y)_0)}{} \end{bmatrix} \\
    &\quad+ (1-\lambda)\begin{bmatrix} \Exp{d(X, g_2(U_2, Y)_1)}{}\\ \Exp{d'(X, g_2(U_2, Y)_0)}{} \end{bmatrix}\\
    &= \lambda \begin{bmatrix} \Exp{d(X, \hat{X}_1)}{}\\ \Exp{d'(X, \hat{S}_1)}{} \end{bmatrix} + (1-\lambda)\begin{bmatrix} \Exp{d(X, \hat{X}_2)}{}\\ \Exp{d'(X, \hat{S}_2)}{} \end{bmatrix}\\
    &=\lambda \begin{bmatrix} D_1\\ D_{s1} \end{bmatrix} + (1-\lambda)\begin{bmatrix} D_2\\ D_{s2} \end{bmatrix}
\end{align}
Now, consider the mutual information terms. We have:
\begin{align}
    &I(X; U_3) - I(U_3; Y)\\
    &=H(X) - H(X|U_3) - H(Y) + H(Y |U_3)\\
    & = H(X) - \lambda H(X|U_1) - (1- \lambda )H(X|U_2) - H(Y) \notag\\
    &\quad+ \lambda H(Y |U_1) + (1-\lambda) H(Y|U_2)\\
    &= \lambda(I(X; U_1) - I(U_1; Y)) + (1-\lambda)I(X; U_2) - I(U_2; Y),
\end{align}
and
\begin{align}
    &\quad R_{WZ}(D_3, D_{s3}) \notag\\
    &= \min_{
    \begin{array}{c} P_{U|X}: \\ \Exp{ {d(X,\hat X)} }{} \leq {D_3} \\\  \Exp{ {d'(X,\hat S)} }{} \leq D_{s3}
    \end{array}} I(X; U) - I(U; Y)\\
    &\leq I(X; U_3) - I(U_3; Y)\\
    &=\lambda R_{WZ}(D_1, D_{s1}) + (1-\lambda)R_{WZ}(D_2, D_{s2}).
\end{align}
This completes the proof of convexity.

Then we provide the proof for Lemma~\ref{lemma_bound_swz}.
The lower bound is achieved when $(\hat{S}, Y)-(\hat{X}, Y) - (U, Y)$ (or $(\hat{X}, Y)-(\hat{S}, Y) - (U, Y)$) forms a Markov chain. In this case, successive refinement~\cite{equitz1991successive} is optimal and the coding rate is determined by the more dominant constraint between $d(X, \hat{X}) \leq D$ and $d_s(S, \hat{S}) \leq D_s$. 
In addition, since $R_{WZ}(D, D_s)$ is convex in both $D$ and $D_s$, the upper bound is achieved by  encoding $X$ and $S$ separately with rates $R_{WZ}(D, \infty)$ and $R_{WZ}(\infty, D_s)$, respectively.

\section{Proof of theorem~\ref{thm_second_coding_rate}}
\label{apx:second_order_coding_rate}
The proof is based on a Gaussian approximation using the following multi-dimensional Berry-Esséen theorem. 
\begin{theorem}[Multidimensional Berry-Esséen theorem\cite{Gotze_clt}]
Let $\vec{U}_1, \vec{U}_2, ... , \vec{U}_n$ be independent random vectors in $\mathbb{R}^k$ with zero mean. Let $\vec{S}_n = \frac{1}{\sqrt{n}}(\vec{U}_1+ ... + \vec{U}_n)$ and $\Cov{\vec{S}_n}{} = \mat{V} > \vec{0}$. Consider a Gaussian random vector $\vec{B}\sim \mathcal{N}(\vec{0}, \mat{V})$, then for all $n\in \mathbb{N}$, we have
\begin{align}
    \sup_{C\in \mathscr{C}_k} |\prob{C}{\vec{S}_n} - \prob{C}{\vec{B}_n}| \leq O\left(\frac{1}{\sqrt{n}}\right)
\end{align}
where $\mathscr{C}_k$ is the family of all convex Borel measurable subsets of $\mathbb{R}^k$.
\end{theorem}
Consider the following information-density-distortion vector
\begin{align}
    \vec{j}(U_i, X_i, Y_i, S_i) = \begin{bmatrix}\iota(X_i, U_i) -\iota(U_i, Y_i)\\ d(X_i, \hat{x}(U_i, Y_i))\\ d_s(S_i, \hat{s}(U_i, Y_i)) \end{bmatrix}
\end{align}
and its first moment 
\begin{align}
    \vec{J} = \begin{bmatrix}I(U; X) -I(U; Y)\\ \Exp{d(X, \hat{X})}{}\\ \Exp{d_s(S, \hat{S})}{} \end{bmatrix}.
\end{align}

Let $\gamma = \log M - \log n$\, using \eqref{eq_excess_bound} allows us to show that there exists a code such that 
\begin{align}
    &\quad \prob{(d_{n,s}(\vec{S}, \hat{\vec{S}}) \geq D_s) \cup
        (d_n(\vec{X}, \hat{\vec{X}}) \geq D) }{} \\
    &\le \mathbb{P}\left[\iota(U;X) - \iota(U;Y) > \gamma \cup  d_s(S, \hat{S}) > D_s\right.\notag\\
    &\qquad \ \left. \cup\  d(X, \hat{X}) > D \right] + \frac{1}{n}\\
    &\leq \prob{\sum_i^{n}\begin{bmatrix}\iota(X_i, U_i) -\iota(U_i, Y_i)\\ d(X_i, \hat{x}(U_i, Y_i))\\ d_s(S_i, \hat{s}(U_i, Y_i)) \end{bmatrix} \geq \begin{bmatrix} \log M \\ nD_s \\ nD \end{bmatrix} - \vec{\log n}}{}\notag\\
    &\quad + \frac{1}{n}\\
    &=\prob{\sum_i^n(\vec{j}_i - \vec{J}) \geq \begin{bmatrix} \log M \\ nD_s \\ nD \end{bmatrix} - n\vec{J} - \vec{\log n}}{} + \frac{1}{n}.
\end{align}
Let 
\begin{align}
    \vec{b^\star} = \sqrt{n}\left [\begin{bmatrix} \frac{1}{n} \log M \\ D \\ G \end{bmatrix} - \vec{J}(\vec{Z}^\star) - \vec{\frac{2\log n}{n}} \right ] ,
\end{align}
where $\vec{\frac{\log n}{n}}$ (and $\vec{\log n}$) denotes the vector $\frac{\log n}{n} \vec {1}_{3}$ ($\log n \vec {1}_{3}$, respectively) with same size as vector $\vec{J}$.
To prove Theorem~\ref{thm_second_coding_rate}, it suffices to show that 
\begin{align}
    \vec{b^\star} \in \mathscr{S}(\mat{V}, \epsilon).
\end{align}

We have 
\begin{align}
    &\quad 1 - \prob{(d_{n,s}(\vec{S}, \hat{\vec{S}}) \geq D_s) \cup
        (d_n(\vec{X}, \hat{\vec{X}}) \geq D) }{} \\
    &\geq \prob{\frac{1}{\sqrt{n}}\sum_i^n \left[\vec{j}_i - \vec{J} \right ]
    \leq 
    \vec{b^\star} + \vec{\frac{\log n}{\sqrt{n}}}}{SUXY} 
     - \frac{1}{n}\\
     &\geq \prob{\vec{B}\leq \vec{b^\star}}{\vec{B}} + O\left(\frac{\log n}{\sqrt{n}}\right)\\
     &\geq 1-\varepsilon,
\end{align}
for n sufficiently large. This completes the proof.
\end{document}